\documentclass[12pt]{article}

\usepackage[dvips]{graphicx}

\usepackage{epsfig,cite}

\usepackage {amsmath}

\usepackage{color}

\usepackage{amssymb}

\usepackage{slashed}

\include{epsf}

\newcount\timecount

\newcount\hours \newcount\minutes  \newcount\temp \newcount\pmhours

\hours = \time

\divide\hours by 60

\temp = \hours

\multiply\temp by 60

\minutes = \time

\advance\minutes by -\temp

\def\hour{\the\hours}

\def\minute{\ifnum\minutes<10 0\the\minutes

            \else\the\minutes\fi}

\def\clock{

\ifnum\hours=0 12:\minute\ AM

\else\ifnum\hours<12 \hour:\minute\ AM

      \else\ifnum\hours=12 12:\minute\ PM

            \else\ifnum\hours>12

                 \pmhours=\hours

                 \advance\pmhours by -12

                 \the\pmhours:\minute\ PM

                 \fi

            \fi

      \fi

\fi

}

\def\monthname{\relax\ifcase\month 0/\or January\or February\or

   March\or April\or May\or June\or July\or August\or September\or

   October\or November\or December\else\number\month/\fi}

\def\bold#1{\setbox0=\hbox{$#1$}%

     \kern-.025em\copy0\kern-\wd0

     \kern.05em\copy0\kern-\wd0

     \kern-.025em\raise.0433em\box0 }

\def\beq{\begin{equation}}

\def\eeq{\end{equation}}

\def\ga{\mathrel{\raise.3ex\hbox{$>$\kern-.75em\lower1ex\hbox{$\sim$}}}}

\def\la{\mathrel{\raise.3ex\hbox{$<$\kern-.75em\lower1ex\hbox{$\sim$}}}}

\def\gev{{\rm \, Ge\kern-0.125em V}}

\def\tev{{\rm \, Te\kern-0.125em V}}

\def\gyr{{\rm \, G\kern-0.125em yr}}

\def\gappeq{\mathrel{\rlap {\raise.5ex\hbox{$>$}}

{\lower.5ex\hbox{$\sim$}}}}

\def\lappeq{\mathrel{\rlap{\raise.5ex\hbox{$<$}}

{\lower.5ex\hbox{$\sim$}}}}

\def\Toprel#1\over#2{\mathrel{\mathop{#2}\limits^{#1}}}

\def\m12{m_{1\!/2}}

\def\bea{\begin{eqnarray}}

\def\eea{\end{eqnarray}}

\def\beqar{\begin{eqnarray}}

\def\eeqar{\end{eqnarray}}

\def\V{{\cal V}}

\def\m{{\cal m}}

\def\B{{\cal B}}

\def\N{{\cal N}}

\begin{document}

\begin{titlepage}

\pagestyle{empty}

\baselineskip=21pt

%\rightline{\tt astro-ph/yymmnnn}

\rightline{KCL-PH-TH/2011-38, LCTS/2011-20, CERN-PH-TH/2011-283}

\rightline{UMN--TH--3019/11, FTPI--MINN--11/27}

\vskip 0.1in

\begin{center}

{\large {\bf Phenomenology and Cosmology of an Electroweak Pseudo-Dilaton and Electroweak Baryons}}

\end{center}

\begin{center}

%\vskip 0.2in

  {\bf Bruce~A.~Campbell}$^{1,2}$,
 {\bf John~Ellis}$^{3,2}$
and {\bf Keith~A.~Olive}$^4$

\vskip 0.1in

{\small {\it

$^1${Department of Physics, Carleton University, \\
1125 Colonel By Drive, Ottawa, Ontario K1S 5B6,
Canada} \\

$^2${TH Division, Physics Department, CERN, CH-1211 Geneva 23, Switzerland}\\

$^3${Theoretical Particle Physics and Cosmology Group, Physics Department, \\
King's College London, London WC2R 2LS, UK}\\

$^4${William I. Fine Theoretical Physics Institute, School of Physics and Astronomy, \\
University of Minnesota, Minneapolis, MN 55455, USA}\\
}}

\vskip 0.2in

{\bf Abstract}

\end{center}

\baselineskip=18pt \noindent

%%%%%%%%%%%%%%%%%%%%%%%%%%%%%%%%%%%%%%%%%%%%%%%%%

{\small

In many strongly-interacting models of electroweak symmetry breaking the lowest-lying
observable particle is a pseudo-Goldstone boson of approximate scale symmetry,
the pseudo-dilaton. Its interactions with Standard Model particles can be described using
a low-energy effective nonlinear chiral Lagrangian supplemented by terms that restore approximate scale symmetry,
yielding couplings of the pseudo-dilaton that differ from those of a Standard Model Higgs boson  by fixed factors. 
We review the experimental constraints on such a pseudo-dilaton in light of new data from the LHC and elsewhere.
The effective nonlinear chiral Lagrangian has Skyrmion solutions that may be identified with
the `electroweak baryons' of the underlying strongly-interacting theory, whose nature
may be revealed by the properties of the Skyrmions.
We discuss the finite-temperature electroweak phase transition in the low-energy effective theory,
finding that the possibility of a first-order electroweak phase transition is resurrected. 
We discuss the evolution of the Universe during this transition and derive an order-of-magnitude lower limit on
the abundance of electroweak baryons in the absence of a cosmological asymmetry, which
suggests that such an asymmetry would be necessary if the electroweak baryons are to provide
the cosmological density of dark matter. We revisit estimates of the corresponding spin-independent
dark matter scattering cross section, with a view to direct detection experiments.
}

%%%%%%%%%%%%%%%%%%%%%%%%%%%%%%%%%%%%%%%%%%%%%%%%

\vfill

\leftline{%CERN-PH-TH/2011-xxx, KCL-PH-TH/2011-xxx, LCTS-2011-yy, 
November 2011}

\end{titlepage}

\baselineskip=18pt

%%%%%%%%%%%%%%%%%%%%%%%%%%%%%%%%%%%%%%%%%%%%%%%%%%

\section{Introduction}

%{\it 

The LHC has demonstrated the need for new physics in the electroweak symmetry-breaking sector.

%}

A couple of inverse femtobarns of data recorded by each of CMS and ATLAS have sufficed to
exclude a Standard Model Higgs boson at the 90\% CL or more between $\sim 130$ and $\sim 500$~GeV~\cite{LHCH}. 
As is well known, if $m_H > 500$~GeV,
the electroweak symmetry-breaking sector becomes strongly-interacting at an energy scale $< 1$~TeV~\cite{HeavyH},
implying the emergence of new physics before that energy, which would also be required for
compatibility with precision electroweak data~\cite{LEPEWWG,Gfitter}. As is also well known, if $m_H < 130$~GeV,
the electroweak vacuum of the Standard Model is destabilized by loop corrections due to the top quark~\cite{lightH}.
It might be argued that one could live with an electroweak vacuum that is metastable on a cosmological
time scale, but most people might feel that this instability should be countered by the intervention of
new physics, though its energy scale is rather uncertain. One way to stabilize the
electroweak vacuum in the light Higgs case is low-energy supersymmetry~\cite{ER}, though other mechanisms are possible.
In addition to these low- and high-mass Standard Model Higgs cases, however, the LHC data are compatible
with a third option for new physics, namely the appearance of a scalar Higgs-like state with
suppressed couplings to Standard Model particles~\cite{suppressed}~\footnote{Or even, Heaven forfend, the absence of
any Higgs-like state at all, an even more exciting scenario for new physics.}. 

This paper is devoted to studying
possible new physics in one generic scenario for such a third way, namely new, nearly conformal
strong dynamics with a (relatively) light pseudo-dilaton that has (many) couplings proportional to
those of a Standard Model Higgs boson, but potentially suppressed by some unknown factor~\cite{suppressed}~\footnote{Other
scenarios are frequently considered, for example that the Higgs boson is a composite Nambu-Goldstone boson arising from
the spontaneous breakdown of some higher-order chiral symmetry~\cite{Contino}.}. It is interesting to note that the
most  ($\sim 1.6 \sigma$) significant fluctuation in the ATLAS/CMS Higgs combination is at $m_H \sim 146$~GeV with a suppressed strength
$\sim$ half the expected signal in the Standard Model~\cite{LHCH}.
Such a `less-Higgs' scenario is not renormalizable, and requires some model-dependent ultra-violet
completion and hence new physics at the TeV scale. Our aim is to explore how much can be said about
such a scenario without reliance on a specific model, and what information low-energy physics might be 
able to provide about this unknown ultraviolet completion.

A general framework for describing such a scenario is provide by 
nonlinear phenomenological Lagrangians. These were introduced in the 1960s
to describe the low-energy interactions of pions~\cite{pionL}, and their extension to include the pseudo-dilaton
of approximate scale invariance of the strong interactions was developed shortly 
thereafter~\cite{dilatonL}. A decade later, the nonlinear chiral Lagrangian formalism was adapted to
describe Higgsless models of electroweak symmetry breaking~\cite{AB}. This formalism has
attracted renewed attention within the last decade, particularly in the framework of
strongly-interacting models of electroweak symmetry breaking~\cite{TC}. Many of these
feature a parametrically light pseudo-Goldstone boson of approximate scale invariance
of the new high-scale strong interactions~\cite{pseudoD,GrinUtt}. The use of a phenomenological
Lagrangian to describe the interactions and other properties 
of such an electroweak pseudo-dilaton~\cite{suppressed} could be justified more
readily than was ever the case for its conjectural counterpart in the
conventional strong interactions, for which parametrical lightness could not be 
demonstrated and no convincing experimental candidate was ever found.

That said, the use of a phenomenological Lagrangian for approximate scale
invariance of the strong interactions provided some useful insights, including the
likelihood that the pseudo-dilaton would decay rapidly into pions, making it
difficult to identify~\cite{pipi}, and the discovery of the canonical trace anomaly~\cite{gammagamma} that
foreshadowed that of QCD and was the forerunner of later calculations of the
Higgs decay rates into $\gamma \gamma$ and gluon pairs~\cite{EGN}. Some twenty years
ago, we also used the phenomenological Lagrangian for approximate scale
invariance to gain insight into the transition between quark-gluon and hadronic
descriptions of QCD~\cite{CEO1,CEO2}, arguing that the chiral transition might be first order and
using the Skyrme description ~\cite{Skyrme} of baryons as chiral solitons to argue that this
transition should be identified with deconfinement.

In this paper, we recycle this approach to explore the phenomenology of scenarios with an
electroweak pseudo-dilaton in generic approximately scale-invariant 
strongly-interacting models of electroweak symmetry breaking. Quite generally, the effective nonlinear
phenomenological electroweak Lagrangian has Skyrmion solutions~\cite{Skyrme}, which 
may be interpreted as `electroweak baryons', whose possible properties we discuss below. These properties could
cast light on the new, high-scale strong dynamics, much as the quantum numbers of conventional
baryons and the rate for $\pi^0 \to 2 \gamma$ decay provided advance evidence for the colour
degree of freedom in QCD. We also use the
phenomenological Lagrangian for approximate scale invariance to gain insight into
the nature of the finite-temperature electroweak phase transition in such a theory~\cite{CJS},
which corresponds to `pseudo-confinement'. We argue that this transition is likely to be
first-order with substantial supercooling and subsequent percolation of bubbles of the true electroweak
vacuum. Finally, exploiting this discussion of the electroweak phase transition, we revisit estimates of the possible 
cosmological relic abundance of such electroweak Skyrmion `electroweak baryons'~\cite{CW}, and discuss
their possible suitability as dark matter candidates~\cite{Nussinov}. We also revisit estimates of the spin-independent cross
section for electroweak baryon scattering on conventional matter~\cite{BDT}, with a view to direct dark matter detection
experiments.

%%%%%%%%%%%%%%%%%%%%%%%%%%%%%%%%%%%%%%%%%%%%%%%%%%

%%%%%%%%%%%%%%%%%%%%%%%%%%%%%%%%%%%%%%%%%%%%%%%%%%

\section{The Nonlinear Effective Electroweak and Dilaton Lagrangian}

As is well known, the tree-level couplings of the Higgs boson of the minimal Standard Model to quarks and leptons
(the bosons $W^+ W^-$ and $Z^0 Z^0$) are directly proportional to their masses (squared), as would be the
case for a dilaton, the conjectural Goldstone boson of spontaneously-broken scale invariance.
In fact, the scale invariance of the Standard Model is broken explicitly by the coefficient of the quadratic term in
the effective potential of the Standard Model, as well as by loop effects. The Higgs sector of the Standard Model
is actually identical with the SU(2) linear $\sigma$ model~\cite{GML}, which was used as a prototype for spontaneous
chiral symmetry breaking in what was subsequently discovered to be QCD~\footnote{Just as the approximate chiral symmetry of the
strong interactions was a first clue that they might be described by a vector-gluon (gauge) theory~\cite{GMOR},
this analogy makes it tempting to conclude that the new, high-scale strong dynamics should also be based on a gauge theory.}.

However, the linear $\sigma$
model  is not the most general phenomenological low-energy Lagrangian for spontaneously-broken chiral
symmetry, a r\^ole that is played by a non-linear chiral Lagrangian~\cite{pionL}, in which there is no candidate for
a dilaton and scale invariance is broken explicitly by the pion decay constant, $f_\pi$. However,
approximate scale invariance can be introduced into the non-linear chiral Lagrangian by
introducing a dilaton $\chi$ whose v.e.v. $V$ breaks scale invariance spontaneously~\cite{dilatonL}. The resulting
non-linear Lagrangian is equivalent to the linear $\sigma$ model if $f_\pi = V$, but this is not the
case in general, and the non-linear model provides a natural one-parameter generalization of the 
linear one, albeit a generalization that is non-renormalizable if $f_\pi \ne V$. 

In a similar way, one
may replace the Higgs sector of the Standard Model by a non-linear Lagrangian for the Goldstone
bosons that are eaten by the $W^\pm$ and $Z^0$, in which the r\^ole of $f_\pi$ is taken by the
electroweak scale $v = 246$~GeV but there is no scalar Higgs boson~\cite{AB}. As in the case of the chiral
Lagrangian, scale invariance may be (approximately) restored by introducing an electroweak
(pseudo-)dilaton field $\chi$~\cite{suppressed}. As we review below, at the tree level its couplings to Standard Model
particles are similar to those of a conventional Higgs boson, but rescaled by a factor $v/V$.
Explicit scale invariance breaking cannot be avoided, due to breaking by loop effects, but can be parametrized
simply if this breaking is small, as might be the case in models with small breaking of scale
symmetry such as walking technicolour~\footnote{In
the case of QCD, scale invariance is broken strongly by the trace anomaly associated with renormalization.}. 
The resulting non-linear phenomenological Lagrangian
provides a minimal generalization of the Standard Model Higgs sector within which, e.g., the
results of searches for the Higgs boson may be interpreted more broadly.

For nonlinear realization of conformal symmetry, in the limit of small explicit breaking,
the nonlinear SU(2) $\times$ SU(2) $\to$ SU(2) effective Lagrangian for the
electroweak Goldstone bosons and the pseudo-dilaton may be written in the form
\begin{eqnarray}
{\cal L} \; & = & \; \frac{v^2}{4} (D_\mu U)(D^\mu U^\dagger) \left( \frac{\chi}{V} \right)^2 + \frac{1}{2} \partial_\mu \chi \partial^\mu \chi 
\nonumber \\
& - & {\cal V}(\chi) - \Sigma_f m_f ( {\bar f}_L f_R U + ~{\rm h.c.}~ )\left(\frac{\chi}{V}\right) + \dots ,
\label{nonlinear}
\end{eqnarray}
where $v = 246$~GeV has the same value as the conventional Higgs v.e.v., $U$ is a $2 \times 2$ unitary
matrix: $U U^\dagger = 1$ that can be parametrized by 3 real degrees of freedom $\pi_i$, which can be identified
with the Goldstone bosons `eaten' by the $W^\pm$ and $Z^0$ bosons, $D_\mu$ is the conventional
electroweak covariant derivative, $\chi$ is the dilaton field and $V$ is its v.e.v. The $f$ are Standard Model fermions (quarks and leptons), 
and the dots indicate higher-order terms in the effective Lagrangian, some of which we discuss in the following. It is apparent from (\ref{nonlinear}) that the couplings of the pseudo-dilaton to gauge bosons and Standard Model fermions 
are related to those of the Standard Model Higgs boson by a simple multiplicative factor $v/V$. 

There are, as well, 
a mass term and self-couplings for the dilaton $\chi$, encoded in the effective potential ${\cal V}(\chi)$. 
These are induced by renormalization and any other (small) explicit breaking of conformal invariance
in the underlying dynamics that gives rise to the effective theory described by (\ref{nonlinear}). 
Small explicit breaking may arise via unsuppressed operators that are almost marginal,
or via operators that are far from marginality, but which enter the dynamics with small coefficients. 
We analyze each of these cases below. 

\subsection{Minimal Violation of Conformal Symmetry}

In the case where the operator $O$ generating explicit breaking of the conformal invariance is almost marginal (ie. when the dimension 
$\Delta_{O}$ of $O$ satisfies $|\Delta_{O}-4| \ll 1$), then ${\cal V}(\chi)$ is calculable to leading order in $|\Delta_{O}-4|$:
\begin{equation}
{\cal V}(\chi) \; = \; B \chi^4 \left[ {\rm ln}(\chi/V) - \frac{1}{4} \right] .
\label{minimal}
\end{equation}
The logarithmic form of the dilaton effective potential is related to the trace anomaly found in~\cite{Schechter,EL},
and the coefficient $B$ is related to the pseudo-dilaton mass:
\begin{equation}
m^2 \; = \; 4 B V^2 .
\label{dilatonmass}
\end{equation}

\subsection{Non-Zero Anomalous Dimensions}

In the case where 
the dilaton effective potential is induced by an operator in the electroweak symmetry-breaking sector which 
has a non-zero anomalous dimension $\gamma$, the effective
potential in (\ref{minimal}) acquires corrections that are ${\cal O}(\gamma)$~\cite{suppressed}. Even if the
anomalous dimension $\gamma$ were not small, conformal symmetry would be approximate if
the coefficient of the anomalous potential term were small, corresponding to a light pseudo-dilaton,
in which case the effective potential would take the form~\cite{GrinUtt}:
\begin{equation}
{\cal V}(\chi) \; = \; \frac{m^2\chi^{4}}{\gamma V^2} \left[ \frac{1}{4 + \gamma} \left(\frac{\chi}{V}\right)^\gamma - \frac{1}{4} \right]
+ {\cal O}\left( \left(\frac{m^2}{V^2}\right)^2 \right) .
\label{gamma}
\end{equation}
Defining ${\hat \chi} \equiv \chi - V$ and expanding (\ref{gamma}) around $V$, we find the following expressions for
the trilinear and quadrilinear ${\hat \chi}$ self-couplings:
\begin{eqnarray}
g_{3 {\hat \chi}} & = & (5 + \gamma + \dots) \frac{m^2}{V} \label{3chi} ,\\
g_{4 {\hat \chi}} & = & (11 + 6 \gamma + \gamma^2 + \dots) \frac{m^2}{V^2} .
\label{4chi}
\end{eqnarray}
Even in the case of violation of conformal symmetry by (almost) marginal operators with $\gamma \to 0$, discussed above, these couplings differ from those in the Standard Model by
significant factors that are in principle measurable in future experiments, such as the high-luminosity LHC~\cite{HiLumLHC}
or CLIC~\cite{CLIC}.

Moreover, pseudo-dilaton/Higgs proportionality would also be violated if there are additional ${\bar f} f$ couplings with
non-zero anomalous dimensions $\gamma_f$, which would be proportional to $(\chi/V)^{1 + \gamma_f}$.
In the absence of a theory of flavour, it is not apparent why such couplings should be flavour-diagonal in
the same basis as the Yukawa couplings giving rise to the fermion masses $m_f$. On the other hand,
any such couplings are constrained by upper limits on flavour-changing neutral interactions, so we
assume here that they are negligible.

\subsection{Anomalous Loop-Induced Couplings}

Another possible deviation from the $v/V$ proportionality rule concerns pseudo-dilaton couplings
to massless gauge bosons $G$, namely gluons $g$ and photons $\gamma$. As is well known, the corresponding 
Standard Model Higgs couplings are generated by
anomalous fermion triangle diagrams (and related diagrams for bosons) that vanish when they have small
masses compared to $m_H$~\cite{EGN}, and are suppressed by the universal $v/V$ factor in the pseudo-dilaton case: 
\begin{equation}
{\cal L}_{GG} \; = \; \left[ \frac{\alpha_s}{8 \pi} b_s 
G_{a \mu \nu} G_a^{\mu \nu} + \frac{\alpha_{em}}{8 \pi} b_{em} F_{\mu \nu} F^{\mu \nu} \right] \left(\frac{\chi}{V}\right) .
\label{triangles}
\end{equation}
However, any additional charged or coloured states that might appear in a
strongly-interacting dynamical theory underlying the pseudo-dilaton and the electroweak Goldstone
bosons would alter the anomaly coefficients $b_{s, em}$~\cite{suppressed}, 
leading to an enhancement in the case of the pseudo-dilaton-$gg$
coupling, or likely a suppression in the case of the pseudo-dilaton-$\gamma \gamma$ coupling, due to a
partial cancellation with the Standard Model $W^+ W^-$ loops.

\begin{figure}[htb]
\begin{center}
\epsfig{file=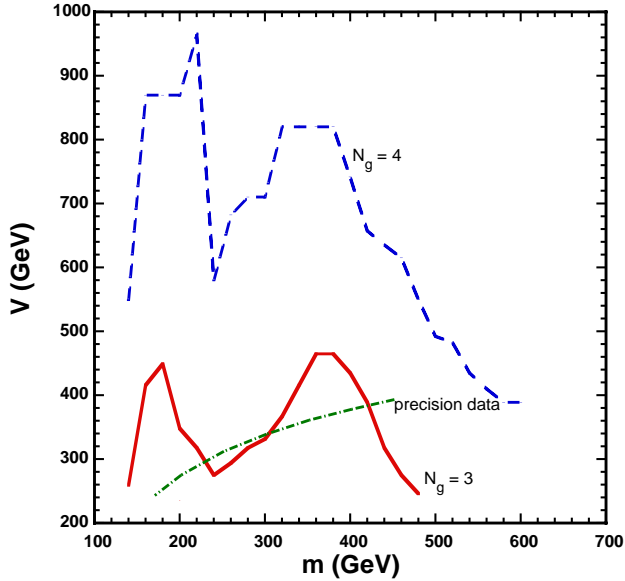,width=0.5\textwidth}
\caption{
\it Lower limits on the pseudo-dilaton v.e.v. $V$ as functions of the its mass $m$,
as obtained from an illustrative combination of the 95\% ATLAS and CMS upper limits on
a Standard Model-like Higgs boson relative to the Standard Model with three generations
(solid red line), and as obtained from the more sensitive of the ATLAS and CMS limits on
the Standard Model with four generations (dashed blue line)~\protect\cite{LHCH}.
Also shown, as a green dash-dotted line, is the lower limit on $V$ as a function of $m$
provided by precision electroweak data, estimated using (\protect\ref{ST}) and the results of~\protect\cite{Gfitter}.
}
\label{fig:LHCH}
\end{center}
\end{figure}

The coefficient $b_s$ in (\ref{triangles}) is normalized so that in the limit $m_H \ll m_t$ the one-loop contribution
from the top quark $\to 2/3$ in the Standard Model, and this a good approximation for $m_{H/\chi} \la m_t$. 
The CMS and ATLAS upper limits~\cite{LHCH} on the production of a Standard Model-like
Higgs boson relative to the Standard Model prediction with the conventional three generations
can be rephrased as an upper limit on $v/V$, and hence a lower limit on $V$ as a function
of the dilaton mass $m$, as indicated by
the solid red line in Fig.~\ref{fig:LHCH}, where we see that the LHC already imposes the limit
$V \ga 400$~GeV in some ranges of $m$. We recall that the most significant fluctuation in the ATLAS/CMS
combination is a $1.6 \sigma$ fluctuation at $m_H \sim 146$~GeV, with a strength $\sim 1/2$ that expected in the
Standard Model~\cite{LHCH}.

Any additional heavy quark would also contribute $\Delta b_s = 2/3$, so that a fourth-generation doublet would
approximately triple the value
of $b_s$ and hence multiply the cross section for $gg \to H$ or $\chi$ by a factor of about 9. However, this enhancement
would be reduced for $m_t \la m_{H/\chi} \la 1000$~GeV, since the top quark contribution yields a contribution to $|b_s| > 2/3$ for this
range of $m_{H/\chi}$~\footnote{We recall that the imaginary part of $b_s \ne 0$ for $m_{H/\chi} > 2 m_t$.}.
On the other hand, since $b_s \to 0$ as $m_{t}/m_{H/\chi} \to 0$,
the top quark contribution to $b_s$ will decrease to $|b_s| < 2/3$ for $m_{H/\chi} \gg m_t$ so that if $m_t \ll m_{H/\chi} \la 1000$~GeV
the enhancement factor due to a fourth generation could exceed 9, depending on the masses of the fourth-generation quarks. 
The searches for a Standard Model-like
Higgs boson at the LHC has already excluded a naive four-generation extension of the Standard Model over a large
range of Higgs masses~\cite{LHCH}. However, a four-generation scenario could be revived in a pseudo-dilaton theory, since the
potential enhancement in $\sigma (gg \to H)$ could be compensated by the universal $(v/V)^2$ factor in the production rate.
Fig.~\ref{fig:LHCH} displays as a dashed blue line the 95\% CL lower limit on $V$ obtained from the CMS and ATLAS
upper limits on a four-generation extension of the Standard Model given in~\cite{LHCH}. For any given value of the
dilaton mass $m$, we use the more sensitive of the ATLAS and CMS limits: this is provided by CMS for $m \le 400$~GeV
and by ATLAS for $m > 400$~GeV. We see that in the four-generation case the LHC already imposes the limit
$V \ga 800$~GeV in some ranges of $m$, and approaches 1~TeV for $m \sim 220$~GeV.

The coefficient $b_{em}$ in (\ref{triangles}) takes the following value in the Standard Model:
\begin{equation}
b_{em} \; = \; \frac{4}{3} \Sigma_f Q_f^2 - 7 ,
\label{bem}
\end{equation}
in the limit of small $m_{H/\chi}$~\cite{MNRuss},
where the first term in (\ref{bem}) includes a sum over all heavy fermions, with the quark
contributions acquiring a colour counting factor of 3, the second term in (\ref{bem}) is due to the $W^\pm$~\footnote{The 
validity of this calculation has recently been questioned in two papers~\cite{GWW} in which the unitary gauge was
used to arrive at a different result. We recall that individual unitary-gauge Feynman diagrams are quartically
divergent, so that great care must be taken to obtain the correct finite result. The relation of unitary-gauge
and $R_\xi$-gauge calculations was discussed in~\cite{FLS}, where it was pointed out that U-gauge
integrals are formally equal to those in the $R_\xi$ gauge in the limit $\xi \to 0$, but that this equivalence
is valid only if the limit is taken after integration over the Feynman parameters. It is known that, if sufficient
care is not taken with the order of limits or in the definitions of loop momenta, the U gauge may yield
incorrect finite results. We also note that, in general, regularization is necessary even in calculations
such as this that yield a finite result, and that neglecting regularization may yield an incorrect finite
result. The result (\ref{bem}) has been verified in many independent calculations.
Therefore, we do not regard the calculation of~\cite{GWW} as evidence of a problem
with dimensional regularization, which has given many correct results in all sectors of the Standard
Model. The papers~\cite{GWW} also contain incorrect remarks about the decoupling
theorem and the trace anomaly. For recent refutations of the conclusions of~\cite{GWW}, see~\cite{FOS}, where issues
in the treatment of divergent integrals of the Goldstone (longitudinal $W^\pm$) modes and decoupling are discussed. 
J.E. thanks Sasha Belyaev, 
David Broadhurst, Mary K. Gaillard, Dimitri Nanopoulos and Douglas Ross for
discussions on these points.}, and we note that the fermion and $W^\pm$ contributions have opposite signs.
The terms in (\ref{bem}) are enhanced in the neighbourhoods of the thresholds where $m_{H/\chi} \sim 2 m_t, 2 m_W$,
respectively, and decrease for larger $m_{H/\chi}$. As a result, $\Gamma (H/\chi \to \gamma \gamma)$ is strongly 
suppressed for $m_{H/\chi} \sim 2 m_t$. The effect of additional
fermions would be to {\it decrease} the magnitude of $b_{em}$ for small $m_{H/\chi}$, but the magnitude might be {\it increased} for
larger masses, depending on the heavier quark masses. Hence the interpretation of the appearance
(or absence) of a $\gamma \gamma$ signal would be ambiguous in a pseudo-dilaton theory.
On the other hand, the search for a $\gamma \gamma$ signal is currently not important for
$M_H \ga 130$~GeV, the range displayed in Fig.~\ref{fig:LHCH}, and we do not discuss it further here. 

Finally, note that because any extra heavy (as yet undiscovered) coloured fermions, such as new fourth-generation quarks, or coloured states in the ultraviolet completion of the
electroweak symmetry breaking sector,
could produce an enhancement in $\sigma (gg \to H)$ that might 
compensate the universal $(v/V)^2$ suppression of the production rate, the observation of an $H/\chi$ state
at a rate consistent with that of $H$ production in the Standard Model would not, by itself, prove that such a state was the 
Standard Model $H$, rather than a $\chi$. Furthermore, since the couplings of both the $H$ and $\chi$ to massive final states 
(which dominate the decay width) are proportional to mass, the measurement of the relative branching ratios to massive final 
states would not resolve this ambiguity. However, this ambiguity could be resolved in two ways. 

One is by a measurement 
of the decay width of the state. Since in both cases the dominant contributions to the decay width are massive states, 
and the only difference in the $\chi$ couplings to the massive states, as opposed to an $H$, is the universal suppression 
by $v/V$, the decay width of a $\chi$ will, to good approximation, be the same as the Standard Model decay width of 
an $H$ of the same mass, but reduced by a factor  $(v/V)^{2}$. So, if the resonance observed is at a mass at which the 
Standard Model predicts an observable decay width, measurement of the decay width at its predicted value would be 
evidence that the state was indeed a Standard Model Higgs. Conversely, if the decay width is measured to be smaller 
than the Standard Model prediction (or appears unmeasurably small) then one will have established that the state is 
not a Standard Model Higgs. Further, if the state is a pseudo-Dilaton $\chi$, and the decay width can be measured, 
that measurement will determine the ratio $v/V$. For $H/\chi$ states of moderately large masses , one of the dominant 
decay modes is to a $Z^0 Z^0$ final state, and the decay chain $H/\chi \rightarrow Z^0 Z^0 \rightarrow \ell^+ \ell^- \ell^+ \ell^-$ provides 
a powerful technique for the measurement of the decay width, and hence the identification of the state as an $H$ or a $\chi$.

A second way to resolve the $H/\chi$ is by disentangling the $gg$ fusion production mechanism from others, such as
$W^+ W^-$ fusion and/or production in association with a $W^\pm, Z^0$ or ${\bar t} t$ pair. In each of these cases,
the production cross section is suppressed by the same factor $(v/V)^2$. This approach to $H/\chi$ discrimination
would work better for a lighter state, whereas measuring the total decay width is easier for a more massive state.

\subsection{Precision Electroweak Data}

Before leaving this Section, we discuss briefly the potential impact of the constraints imposed
by precision electroweak data via determinations of the $S$ and $T$ vacuum polarization
parameters~\cite{ST}. The contributions of pseudo-dilaton loops are proportional to those from a conventional
Higgs boson, but scaled by factors $(v/V)^2$:
\begin{eqnarray}
\Delta S (\chi) & = & \left( \frac{v}{V} \right)^2 \frac{1}{12 \pi} {\rm ln} \left( \frac{m^2}{m_Z^2} \right) + \dots \\
\Delta T (\chi) & = & - \left( \frac{v}{V} \right)^2 \frac{3}{16 \pi \cos \theta_W^2} \left( \frac{m^2}{m_Z^2} \right) + \dots
\label{ST}
\end{eqnarray}
where the dots denote terms that do not depend logarithmically on the dilaton mass.
Standard Model fits to the precision electroweak data are compatible with a Higgs mass $\sim m_Z$,
and hence a negligible contribution to $S$ and $T$ from the electroweak symmetry-breaking sector.
The potential suppressions of the dilaton contributions by factors $(v/V)^2$ open the possibility that even
a relatively heavy dilaton could be compatible with the precision electroweak data. Specifically, Fig.~10
of~\cite{Gfitter} shows that the pseudo-dilaton contributions to $S$ and $T$ (\ref{ST}) would lie
within the present experimental upper bounds even for $m \sim 1$~TeV, if $(v/V)^2 \sim 1/10$.
We display in Fig.~\ref{fig:LHCH} as a green dash-dotted line the lower limit on $V$ as a function of $m$
obtained using the forms (\ref{ST}) of the loop corrections and the analysis of~\cite{Gfitter}
that found $m_H < 169$~GeV at the 95\% CL in the Standard Model, i.e., when $V = v$, assuming three generations. 
We see that this gives a lower bound on $V$ that is weaker than the direct Higgs searches in the three-generation
case (red line) over much of the range displayed. However, the limit from precision electroweak data is stronger 
than that from the direct Higgs searches in at large $m$, and is competitive in an intermediate range of 
$m$~\footnote{We do not discuss here the evaluation of the precision electroweak limit in the four-generation case,
as it requires some supplementary hypotheses on the spectrum of fourth-generation quarks and leptons.}.

Nevertheless, a full
discussion of the precision electroweak constraints requires a better understanding of the other contributions to $S$ and $T$
in a strongly-interacting theory, which are quite model-dependent. For example, a massive $\rho$-like
vector resonance with self-coupling $g_\rho \epsilon_{abc} \partial_\mu\rho^a_\nu \rho^b_\mu \rho^c_\nu$
would contribute~\cite{rhobit}
\begin{equation}
\Delta S (\rho) \; = \; \frac{4 \pi}{g_\rho^2} ,
\label{rho}
\end{equation}
and other contributions are to be expected in any specific model.
Therefore, it may be premature to conclude that the precision electroweak excludes any region of the
pseudo-dilaton parameter space at the present level of understanding.

%%%%%%%%%%%%%%%%%%%%%%%%%%%%%%%%%%%%%%%%%%%%%%%%%%

%%%%%%%%%%%%%%%%%%%%%%%%%%%%%%%%%%%%%%%%%%%%%%%%%%

\section{Electroweak Baryons}

If the searches for a Higgs boson do indeed discover a pseudo-dilaton, particle physics will be in a
situation reminiscent of, but rather different from, that of strong-interaction physics in the 1960s. The similarity is that
one has an effective low-energy theory but is ignorant
of its ultraviolet completion and short-distance structure. In the 1960s, the approximate chiral symmetry
of the low-energy effective theory motivated the suggestion that the underlying theory might be a gauge theory, and
supplementary information, such as the rate for $\pi^0 \to \gamma \gamma$ decay and the success of the quark model
in describing baryons, in particular, suggested that the gauge group should be the SU(3) of colour, i.e., QCD.
In the present case, we have less information. We have an effective low-energy Lagrangian,
its chiral symmetry suggests that the underlying theory is a gauge theory,
and the LHC is providing constraints on the pseudo-dilaton couplings.
However, we currently have no analogue of the phenomenological success of the quark model to guide
us towards the nature of the dynamics of any underlying theory.

However, as we discuss below,
relevant experimental information may be provided by `electroweak pseudo-baryons', which
arise as Skyrmion solutions of the nonlinear effective electroweak Lagrangian. Their
existence is `inevitable' but their properties are model-dependent, 
and discovering any such states and measuring their properties
would provide valuable insights into the underlying theory~\footnote{For related studies in
`Little Higgs' models, see~\cite{ZH}.}.

It is well known that the nonlinear SU(3) $\times$ SU(3) $\to$ SU(3) and SU(2) $\times$ SU(2) $\to$ SU(2)
chiral Lagrangians of QCD have topological soliton solutions~\cite{Skyrme,Witten} called
Skyrmions, as long as there is a higher-order `Skyrme term':
\begin{equation}
{\cal L_{S}} = \frac{1}{32e^{2}} 
Tr[(D_\mu U)U^\dagger,(D_\nu U)U^\dagger]^{2} .
\label{Skyrme}
\end{equation}
These solitons appear because $\pi_3$(SU(2)) = Z,
and have integer baryon number
\begin{equation}
B = \frac{1}{24 \pi^2} \int d^3x \epsilon^{ijk} Tr\left[(U^{-1}\partial_iU)(U^{-1}\partial_jU)(U^{-1}\partial_kU)\right].
\label{Bnumber}
\end{equation}
In the case of the nonlinear SU(3) $\times$ SU(3) $\to$ SU(3) theory
the effective Lagrangian contains
a Wess-Zumino term $N \Gamma$, where $N$ is an integer, and the lowest-lying
$B = 1$ baryon is a fermion (boson) with $I = J = 1/2$ $(I = J = 0)$ if $N$ is odd (even). If the
underlying strongly-interacting theory is a non-Abelian SU(N) gauge theory, $N$ is
identified with the number of colours, and in QCD the $B = 1$ baryon is necessarily a fermion because $N = 3$.
In the case of the nonlinear SU(2) $\times$ SU(2) $\to$ SU(2) theory,
there is no Wess-Zumino term, and the $B = 1$ baryon is not necessarily a fermion.
Nevertheless, the topology accommodates this possibility, since $\pi_4$(SU(2)) = Z$_2$~\cite{Witten}.

The above discussion can be taken over intact to the case of the scale-invariant electroweak
$\times$ SU(2) $\to$ SU(2) chiral Lagrangian. The only difference is the appearance of a
dilaton-dependent factor in the leading-order term in the chiral Lagrangian (but not the
Skyrme term (\ref{Skyrme})), which does not affect the topological properties of the theory. Indeed, a $B = 1$ Skyrmion solution 
of the field equations for the dilaton-dependent SU(3) $\times$ SU(3) $\to$ SU(3)
chiral QCD Lagrangian has been exhibited in~\cite{EK}, with the same qualitative features as
in the dilaton-free theory. We therefore conclude that the (approximately)
scale-invariant nonlinear electroweak Lagrangian possesses Skyrmion solutions with non-zero
electroweak pseudo-baryon number ${\cal B} \ne 0$, which may be either baryons or fermions, depending on
the nature of the underlying strongly-interacting theory.

As pointed out in~\cite{Witten}, there are other exotic possibilities for the electroweak baryon quantum number,
which should also be considered in the absence of information about the ultraviolet completion of the effective
nonlinear Lagrangian.
For example, if the underlying strongly-interacting theory is based on an SO(N) gauge theory,
no distinction is possible between baryons and antibaryons, i.e., ${\cal B}$ is a Z$_2$ quantum number. In this case,
baryons are produced in pairs and two baryons may annihilate into $N$ mesons. Another, more
exotic possibility arises if the
underlying gauge theory is of Sp(N) type. In this case there are no stable baryons at all, and any wannabe baryon
could decay directly into $N$ mesons.

It has been suggested recently that the ultraviolet completion of the electroweak symmetry-breaking
sector of the Standard Model might be via classical field configurations, dubbed `classicalons'~\cite{classicalons}.
In the case of QCD, the ultraviolet completion of the low-energy chiral theory involves an infinite
series of resonant states, most of which are unstable against hadronic decays. The only states that
are stable against hadronic decays are the lowest-lying baryons. From the point of view of this
discussion, that is because they are the lowest states in a non-trivial topological sector. Indeed,
in QCD these are the only known non-trivial semi-classical field configurations. However, they do not
play an important r\^ole in unitarizing $\pi \pi$ scattering. As discussed above, though, they do encode
key features of the underlying theory. In a similar way, the Skyrmions of the electroweak theory may
encode important aspects of the dynamic underlying electroweak symmetry breaking that
is not determined by the effective low-energy Lagrangian alone. However, they would not provide
a realization of the `classicalon' suggestion.

%%%%%%%%%%%%%%%%%%%%%%%%%%%%%%%%%%%%%%%%%%%%%%%%%%

%%%%%%%%%%%%%%%%%%%%%%%%%%%%%%%%%%%%%%%%%%%%%%%%%%

\section{The Electroweak Phase Transition}

We now discuss the behaviour of the pseudo-dilaton theory at finite temperature, $T$,
focusing in particular on the electroweak phase transition in this model~\footnote{For an early
study of technicolour cosmology, see~\cite{CW}, and for a more recent discussion see~\cite{CJS}.
For a more closely related recent discussion of cosmology in a nearly conformal model, see~\cite{KS}.}. To this
end, we first discuss the $T$ dependences of its three parameters with non-zero
mass dimensions, namely $v, V$ and $m$, whose leading corrections are in general 
of the forms ${\cal O}(T^2/v^2, T^2/V^2)$. 

\subsection{Finite-Temperature Corrections}

Since $v$ appears in the nonlinear
electroweak Lagrangian in the same way as $f_\pi$ in the much-studied nonlinear pion
Lagrangian of QCD, the leading $v$-dependent temperature corrections to $v$ itself are known from studies of that theory~\cite{GL}:
\begin{equation}
v(T, v) \; = \; v \times \left[ 1 - \frac{{\N} T^2}{24 v^2} + {\cal O}\left(\frac{T^4}{v^4}\right) \right] ,
\label{vTv}
\end{equation}
where $\N$ is the number of electroweak `pseudo-flavours'. In the minimal realization of the Standard
Model with an SU(2) $\times$ SU(2) $\to$ SU(2) structure, $\N = 2$, but one could imagine embedding
the theory in a larger structure with more `pseudo-pions', in which case $\N > 2$. In QCD, the leading
$T/v$-dependent corrections to $\langle 0 | {\bar q} q | 0 \rangle$ have also been calculated, and are
of the form $\left[ 1 - (\N^2 - 1) T^2 / (12 \N v^2) + {\cal O}(T^4/v^4) \right]$~\cite{GL}. Numerical evaluations
suggest that $v(T, v)$ and $\langle 0 | {\bar q} q | 0 \rangle$ do not vanish for $T < 2 v$, but the 
$T/v$ expansion is in principle valid only in the limit $T \ll v$, so this conclusion should be taken
with a pinch of salt. Moreover, since the kinetic term for the Goldstone bosons in the
nonlinear effective electroweak Lagrangian also has a factor $\chi^2/V^2$, there is also
a finite-temperature correction to $v$ that are $\propto T^2/(12V^2)$.
The leading contributions to the finite-temperature dilaton effective potential are also well known:
\begin{equation}
\delta \V (T, \chi) \; = \; \frac{B}{12}  \chi^2 \left[{\rm ln}(\frac{\chi}{V}) + \frac{1}{3}\right] T^2 + \dots ,
\label{deltaVT}
\end{equation}
where the dots indicate terms that are of higher order in $T/V$.

\subsection{Critical Temperature}

The finite-temperature correction (\ref{deltaVT}) does not alter the fact that the minimum
of the effective potential is at $\langle \chi \rangle \ne 0$. However, the preferred finite-temperature
vacuum is found by minimizing the free energy
\begin{equation}
F \; \equiv \; - P + \V(\chi) + \delta \V(T, \chi) ,
\label{free}
\end{equation}
where
\begin{equation}
P \; = \; \left( N_B + \frac{7}{8} N_F \right) \frac{\pi^2}{90} T^4
\label{P}
\end{equation}
for $N_B$ massless bosons and $N_F$ massless fermions, and hereafter we denote
${\cal N} \equiv N_B + \frac{7}{8} N_F$. The finite-temperature
electroweak phase transition is driven by the difference between (\ref{P}) in the
high-temperature vacuum with $\langle \chi \rangle = 0$ and the low-temperature
vacuum with $\langle \chi \rangle = V$. 

The contributions of the Standard Model particles
are the same in both states: the difference arises from the numbers of degrees of freedom
in the electroweak symmetry-breaking sector above and below the transition
temperature. In the low-temperature theory, this is simply $N_B = 4$ for the 
pseudo-dilaton and the longitudinal polarization states of the $W^\pm$ and $Z^0$.
If the high-energy theory were a parity-conserving SU(N) gauge theory with $N_f$ multiplets of
fermions in the fundamental representation, it would contribute
\begin{equation}
\Delta {\cal N} \; = \; 2 (N^2 - 1) + \frac{7}{2} \left( N \times N_f \right) .
\label{N1}
\end{equation}
If this theory were to be approximately conformal, one would have $N_f = {\cal O}(11 N/2)$,
and hence
\begin{equation}
\Delta {\cal N} \; \sim \; \frac{85}{4} N^2 - 6 .
\label{N2}
\end{equation}
Substituting into (\ref{P}), it is clear that, for any plausible ${\cal N}$, the difference between the
pressures in the symmetric and broken phases is
\begin{equation}
\Delta P \; = \; \Delta {\cal N} \; \frac{\pi^2}{30} \; T^4 \; \sim \; {\cal O}({\rm few}) \; \pi^2 \; T^4 .
\label{approxP}
\end{equation} 
This is a useful order of magnitude to keep in mind in the following discussion -
the exact numerical coefficient does not play an essential r\^ole.

The critical temperature, $T_c$, is calculated by equating $F$ in the symmetric and
broken phases:
\begin{equation}
\Delta P \; = \; - \V(\chi_T) - \delta \V(T, \chi_T) ,
\label{critical}
\end{equation}
where $\chi_T$ denotes the value of $\chi$ in the broken phase at finite-temperature.
Neglecting $\delta \V(T, \chi)$, so that $\chi_T \sim V$, and using $\V = - (B/4) V^4$ at
the minimum in the broken phase and recalling that $m^2 = 4 B V^2$, one has
\begin{equation}
T_c \; = \; {\cal O}\left(\frac{1}{2\sqrt{\pi}}\right)\sqrt{m V} + \dots \; \simeq \; 0.28 \sqrt{mV} .
\label{Tc}
\end{equation}
At this temperature, $\delta \V(T_c, V) \sim - m^2 T_c^2/12$ and, using the
approximation (\ref{Tc}) for $T_c$, we see that $|\delta \V(T_c, V)| = {\cal O}(m^3 V)
\ll |\V (V)|$, implying that the estimate (\ref{Tc}) is valid to leading order in $m/V$
and that in the same approximation $\chi_{T_c} = V$.

We infer that the electroweak phase transition is likely to have been first-order, at least in the limit of
small dilaton mass $m$, as might have been expected from the `Coleman-Weinberg'
form of the effective dilaton potential $\V(\chi)$ (see also~\cite{KS})~\footnote{{\it A priori}, this re-opens
the possibility of electroweak baryogenesis, a possibility excluded in the Standard
Model by the LEP lower limit on $m_H$. However, realizing this option in practice
would require additional CP violation beyond the Kobayashi-Maskawa phase, an
issue beyond the scope of this work.}.

\subsection{Tunnelling and Supercooling}

Since the phase transition is first-order, it will have proceeded by tunnelling, and
it is to be expected that the early Universe supercooled
to some extent before completing the transition.
The rate for tunnelling between two vacua whose energy densities differ by $\Delta \V$ is
\begin{equation}
\Gamma \; \sim \; \Lambda \times {\rm exp} \left(- \frac{27 \pi^2}{2} \frac{S_0^4}{(\Delta \V)^3} \right) ,
\label{GTW}
\end{equation}
where one expects $\Lambda \sim V$ and the bounce action $S_0$ is given in the thin-wall
approximation by
\begin{equation}
S_0 ~{\rm (thin-wall)}~ \; = \; \int_0^V \sqrt{2 \V} d\chi \; \equiv \alpha \sqrt{B} V^3 .
\label{bounce}
\end{equation}
Integrating naively the logarithmic potential (\ref{nonlinear}) would yield $\alpha = 0.37 \sqrt{2}$,
but this potential is not accurate close to the origin at finite temperatures. If instead one
integrates (\ref{bounce}) from $V/2$ to $V$ and doubles the result, one finds 
$\alpha \sim 0.26 \sqrt{2}$: for numerical estimates we estimate $\alpha \sim 0.4$. The
condition for completing the transition is $\Gamma/H \sim 1$, where $H$ is the
Hubble expansion rate. This condition would be satisfied if there was a temperature
$T < T_c$:
\begin{equation}
{\rm ln}\left( \left( \frac{3}{2 \pi B} \right)^{1/2} \frac{m_P }{V} \right) \; \sim \; \frac{27 \pi^2}{2} \frac{\alpha^4 B^2 V^{12}}{(\Delta {\cal N}\pi^2/30)^3 (T_c^4 - T^4)^3} ,
\label{complete}
\end{equation}
where $\Delta {\cal N}$ here is the difference between the numbers of light particles before and after tunnelling, 
which we estimate to be $\Delta {\cal N} = {\cal O}(100)$ as discussed previously, and we have used the Hubble parameter
corresponding to the vacuum energy density. It is easy to see that the condition (\ref{complete})
cannot be satisfied for representative values of $m/V \sim 1/10 - 1$ corresponding to $B \sim 1/400 -1/4$,
if we use the thin-wall approximation (\ref{bounce}) to evaluate the tunnelling rate (\ref{GTW}). Using the high-temperature
form $(16 \pi/3) (S_0^3/(\Delta \V)^2)$ for the bounce action yields the same conclusion, so we infer that the
transition must occur at some temperature $T \ll T_c$ where the barrier is very low and the thin-wall
approximation (\ref{bounce}) is inapplicable~\cite{KS,RS}.

\begin{figure}[htb]
\begin{center}
\epsfig{file=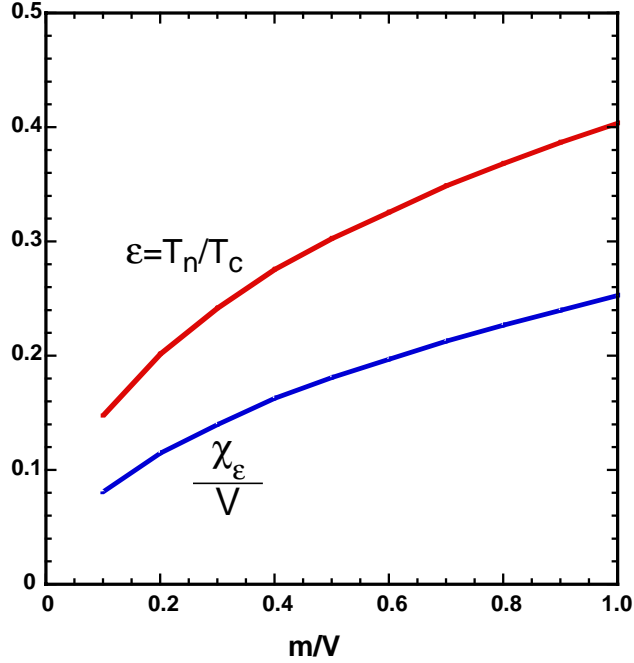,width=0.5\textwidth}
\caption{
\it The ratio of the nucleation temperature $T_n$ relative to the critical temperature $T_c$, 
$\epsilon \equiv T_n/T_c$, and the ratio of the corresponding value of $\chi$ at nucleation,
$\chi_\epsilon \equiv \chi(T_n)$, to the
zero-temperature v.e.v. of the pseudo-dilaton field $V \equiv \langle 0 | \chi | 0 \rangle$,
as functions of the ratio of the dilaton mass $m$ to the zero-temperature v.e.v. of $\chi$, $V$.
}
\label{fig:Tc}
\end{center}
\end{figure}

Therefore, we explore the possibility that the supercooled transition takes place at a nucleation temperature $T_n = \epsilon T_c
\ll T_c$ to a `small' value $\chi_\epsilon \ll V$ of the dilaton field where
\begin{equation}
\V(\chi_\epsilon) \sim - {\cal N} T_n^4 = - {\cal N} \epsilon^4 T_c^4 = - \epsilon^4 \frac{B V^4}{4} ,
\label{epsiloncondition}
\end{equation}
which reduces to solving the condition
\begin{equation}
\left(\frac{\chi_\epsilon}{V}\right)^4 \left[ {\rm ln}(\chi_\epsilon/V) - \frac{1}{4} \right] \; = \; - \frac{\epsilon^4}{4} .
\label{chiepsilon}
\end{equation}
Corresponding to $\chi_\epsilon$, one may estimate 
$\alpha_\epsilon \equiv \int_0^{\chi_\epsilon} \sqrt{2 \V} d \chi/(\sqrt{B} V^3)$,
and then solve the $\Gamma/H \sim 1$ consistency condition, which may be cast in the form
\begin{equation}
\epsilon \; = \; \left[ 1 - \left( \frac{864 \pi^2 \alpha_\epsilon^4}{ B {\rm ln}((3/2\pi B)^{1/2} m_P/V)} \right)^{1/3} \right]^{1/4} .
\label{fixepsilon}
\end{equation}
Using this procedure, for $B \sim 1/400$, corresponding to $m/V = 1/10$, we estimate
\begin{equation}
\epsilon \; = \; \frac{T_n}{T_c} \; = \; 0.15; \; \frac{\chi_\epsilon}{V} \; = \; 0.08; \; \alpha_\epsilon \; = \; 0.04 \sqrt{2} ;
\label{epsilonvalue}
\end{equation}
i.e., the Universe supercools substantially to a temperature $\sim 0.15$ of $T_c$ before completing
a first-order electroweak phase transition to a dilaton v.e.v. $\sim 0.08$ of its present value $V$.
Fig.~\ref{fig:Tc} illustrates the $m$ dependences of $\epsilon \equiv T_n/T_c$ and $\chi_\epsilon/V$,
where $\chi_\epsilon \equiv \chi_{T_n}$. We see that $T_n \ll T_c$, and that also the value of $\chi$ at
nucleation $\chi_\epsilon \ll V$.

In making the above estimate, we have assumed that the Universe is radiation-dominated
for temperatures $T > T_c$. Both before and after nucleation, the scalar field energy density is given
approximately by ${\cal V}(0) = B V^4/4$, and radiation in fact dominates the overall energy density
at temperatures $T > T_{eq}$, where
\begin{equation}
T_{eq} \; \simeq \; \left( \frac{30 {\cal V}(0)}{\pi^2 {\cal N}} \right)^{1/4}  \; = \; \left( \frac{30}{\pi^2 {\cal N}} \right)^{1/4}
\left( \frac{B}{4} \right)^{1/4} \; \simeq \; 0.21 \sqrt{mV} .
\label{Teq}
\end{equation}
The estimate (\ref{Teq}) of $T_{eq}$ is 
somewhat smaller than the estimate (\ref{Tc}) of the critical temperature $T_c$, but rather larger than
the estimate (\ref{epsilonvalue}) of the nucleation temperature $T_n$. This indicates that, for some
period around the time of nucleation, the scalar field energy density dominates somewhat over radiation. 
However, the dilaton field rolls very quickly, with a characteristic
time scale $t_{roll} \sim 200/V$, which is very short compared to the 
nucleation time scale $t_n \sim 10^{18}/V$, and decays with a characteristic
rate that is very large compared with the Hubble expansion rate. Hence the 
Universe does not enter a de Sitter expansion phase (inflationary epoch),
because the conventional slow-roll conditions
\begin{equation}
\frac{1}{16 \pi G_N} \left( \frac{\V^\prime}{\V}\right)^2 \; \ll \; 1 , \; \;
\frac{1}{8 \pi G_N} \left( \frac{\V^{\prime\prime}}{\V}\right) \; \ll \; 1 ,
\label{slowroll}
\end{equation}
where $G_N$ is Newton's constant, are strongly violated.

\begin{figure}[htb]
\begin{center}
\epsfig{file=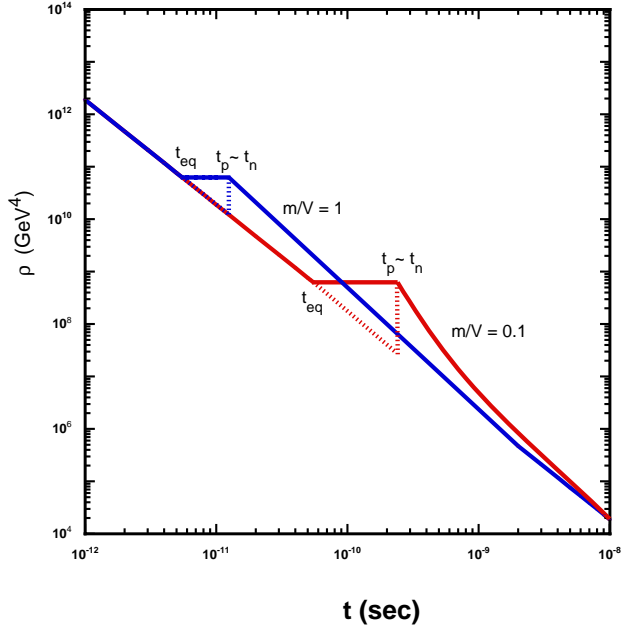,width=0.5\textwidth}
\caption{
\it Illustration of the cosmological evolution of the energy density, $\rho$, for $m/V = 1$ and $0.1$ (blue and red
solid lines, respectively).  At high temperatures $T > T_{eq}$, $\rho$ is
dominated by relativistic particles, including electroweak constituents as well as Standard Model particles.
This is followed by a short epoch during which $\rho$ is dominated by the scalar field energy and the
energy density of relativistic particles drops (dotted lines), which
is terminated by nucleation and percolation of bubbles of the electroweak vacuum. The scalar field energy density is
then converted rapidly to relativistic Standard Model particles.
}
\label{fig:rho}
\end{center}
\end{figure}

The above discussion is summarized in Fig.~\ref{fig:rho}, in which the solid blue (red) lines represent the 
evolution of the cosmological energy density, $\rho$, for $m/V = 1 (0.1)$. Relativistic particles, including electroweak 
constituents (which we estimate to contribute $\Delta {\cal N} \simeq 100$) as well as Standard Model particles
(which contribute $\Delta {\cal N} = 427/4$) dominate $\rho$ until it falls to the temperature $T_{eq}$ at which it
becomes equal to the scalar field energy, which $\simeq {\cal V}(0)$. Following the change in the rate of
evolution of $\rho$, the scalar field energy density dominates
$\rho$ during a short epoch that is terminated by nucleation and (almost immediately)
percolation of bubbles of the low-energy electroweak vacuum. Since the dilaton field rolls and decays
with a rate $\propto \alpha_\chi m$ where $\alpha_\chi$ is some coupling strength,
that is much larger than the Hubble expansion rate, the scalar field energy density is then
converted rapidly and efficiently to relativistic Standard Model particles, as indicated by the later kinks in the solid lines.
The dotted lines represent the evolution of the energy density of relativistic particles during the short
epoch of scalar field energy domination.

The short period of non-adiabatic expansion between $T_{eq}$ and $T_n$ is accompanied by
a corresponding growth in entropy. The growth in the scale factor is by a factor $\simeq R_n/R_{eq} =$ exp$[H (t_n - t_{eq})] = T_{eq}/T_n \sim
1.9/\sqrt{m~[{\rm TeV}]}$ for 1~TeV~$ > m > $~100~GeV, leading to a corresponding growth in the entropy by a
factor of $(R_n/R_{eq})^3 \sim 7$ to 200. This is not a big problem for baryogenesis before the electroweak phase transition, but
this entropy growth should be taken into account, particularly if the dilaton mass $m \sim 100$~GeV.

\subsection{Percolation}

The transition to the new electroweak vacuum would have been completed at the percolation time $t_p$,
when the Universe was filled with bubbles of true vacuum~\cite{GW}. In order to estimate this, we first
recall that the probability for bubble nucleation is
\begin{equation}
{\cal P} \; \sim \; V^4 e^{- {\cal B}} ,
\label{probability}
\end{equation}
where the bounce action
\begin{equation}
{\cal B} \; \sim \; \frac{27 \pi^2}{2} \frac{\alpha^4 B^2 V^{12}}{{\cal N}^3 T_c^{12}} ,
\label{Bounce}
\end{equation}
where ${\cal N} T_c^4 = B V^4/4$. Following nucleation, the fractional volume of space remaining in the
false vacuum is
\begin{equation}
f (t) \; \sim \; {\rm exp}[{- \int^t_{t_n} d \tau {\cal P}(\tau) a(\tau)^3 V(\tau, t)}] ,
\label{fraction}
\end{equation}
where
\begin{equation}
V(\tau, t) \; = \; \frac{4 \pi}{3} \left( \int^t_\tau \frac{d \tau^\prime}{a(\tau^\prime)} \right)^3 ,
\label{V}
\end{equation}
For definiteness, 
we assume vacuum domination, so that $a(t) \sim e^{Ht}$. We find
that the percolation time $t_p = t_n (1 + \delta)$, where the small correction $\delta$
is found by solving the equation
\begin{equation}
- {\rm ln} f \; = \; \frac{\pi}{3} t_n^4 {\cal P} \delta^4 ,
\label{solvef}
\end{equation}
which yields the estimate
\begin{equation}
\delta \; \sim \; \frac{10^4}{V t_n} \; \sim \; \frac{10^4 T_n^2}{V M_P} \; \ll \; 1 .
\label{delta}
\end{equation}
We conclude that the phase transition would have completed almost immediately
after nucleation, with the percolation time $t_p \sim t_n$. This conclusion is independent
of the exact expansion rate between $t_n$ and $t_p$.

\subsection{Confinement}

As was pointed out in~\cite{CEO1,CEO2}, it is a natural possibility to identify the high-temperature dilaton
phase transition with the deconfinement transition. In the case of QCD, the argument
was as follows. In the nonlinear Lagrangian for pseudoscalar mesons with no dilaton,
the Skyrmion mass $\propto f_\pi$, and the Skyrmion radius $\propto 1/f_\pi$. Interpreting
the Skyrmion as a baryon, we interpret the low-mass, large-radius limit as $f_\pi \to 0$ as
the quark deconfinement limit. Finally, since $m_\pi^2 f_\pi^2 = m_q \langle 0 | {\bar q} q | 0 \rangle$ 
in QCD, the chiral transition when $\langle 0 | {\bar q} q | 0 \rangle \to 0$ requires the vanishing of
$f_\pi$ and hence deconfinement in QCD. 

The next question is the relationship of the chiral condensate $\langle 0 | {\bar q} q | 0 \rangle \to 0$
to the dilaton v.e.v. In the nonlinear QCD Lagrangian with a dilaton,
$V = \langle 0 | \chi | 0 \rangle \ne 0$ is in fact required in order that $\langle 0 | {\bar q} q | 0 \rangle \ne 0$~\cite{CEO1,CEO2}.
The explicit
construction of a Skyrmion solution to this theory demonstrated that the baryon mass $ \to 0$
and the baryon radius $\to \infty$ when $V \to 0$. If $\langle 0 | {\bar q} q | 0 \rangle$ vanishes at
the same temperature as the dilaton v.e.v., $V$, the chiral symmetry/deconfinement transition is identified
with spontaneous scale symmetry breaking.

On the other hand, as also pointed out in~\cite{CEO1,CEO2},
it is possible that the chiral finite-temperature corrections to the effective chiral Lagrangian, cf (\ref{vTv}),
drive $\langle 0 | {\bar q} q | 0 \rangle \to 0$ even if $ V = \langle 0 | \chi | 0 \rangle \ne 0$. In this case the
chiral symmetry/confinement transition would be distinct from the dilaton transition, taking place
at a lower temperature. In the case of QCD with two (and one moderately) light flavours, 
it was not possible on the basis of the effective
low-energy theory alone to determine whether finite-temperature chiral corrections drive
$\langle 0 | {\bar q} q | 0 \rangle$ and $f_\pi \to 0$ below the dilaton transition temperature, or not.
However, lattice calculations of QCD seem to favour strong simultaneous variations in the ${\bar q}q$ and
gluon condensates around the quark-gluon/hadron transition, consistent with the coincidence
of chiral symmetry breaking, quark confinement and the development of a dilaton v.e.v.

In the electroweak case, by analogy, the formation of a dilaton v.e.v. $V \ne 0$ is
necessary for the breaking of chiral symmetry that leads to the nonlinear electroweak
effective Lagrangian. Moreover, on the basis of the Skyrmion model for
electroweak baryons we argue that the breaking of electroweak chiral symmetry can be identified with
pseudo-confinement, and expect that the electroweak baryon ${\B}$ has a mass $m_\B = {\cal O}(V)$,
with a numerical coefficient that depends on the underlying strongly-interacting theory.

The remaining question is whether the chiral transition coincides
with the dilaton transition, or whether it occurs at a lower temperature, driven by the
finite-temperature corrections (\ref{vTv}). On the basis of this formula, and more detailed
evaluations of higher-order finite-temperature corrections in the QCD case~\cite{GL}, it seems
reasonable to hypothesize that they would not drive $v \to 0$ at a temperature below $\sim 2 v \sim 500$~GeV.
On the other hand, (\ref{Tc}) suggests that $T_c \sim 100$~GeV, and we saw subsequently that
there is likely to have been supercooling by a factor $\sim 10$ before the transition to $V \ne 0$.
According to the above estimates, the chiral finite-temperature corrections would not drive $v \to 0$ below this temperature, 
so we expect that the transitions to $V \ne 0$ and $v \ne 0$ occurred simultaneously,
and we identify this transition with the pseudo-confinement of whatever underlying constituents
there may have been.

\subsection{Electroweak Baryon-to-Entropy Ratio}

The above discussion of electroweak baryons provides a framework for re-examining the
longstanding suggestion that technibaryons might provide the astrophysical cold dark matter~\cite{Nussinov}~\footnote{For
the record, we note that electroweak baryons are expected to be sufficiently stable to serve as dark matter particles~\cite{DF}.}.
When the electroweak phase transition was completed at the time of percolation by bubbles
filling space, the relative orientations of the chiral condensates 
in the bubble interiors would have been partially mismatched, in general, leading {\it \`a la} Kibble to the appearance
at their boundaries of topologically stable defects~\cite{Kibble}. The number density 
of the defects at the time of percolation would have been of the same order as the bubble density at percolation~\footnote{Since,
as argued earlier, we expect the electroweak phase transition in the class of models studied here
to have been first-order, we do not expect modifications of the Kibble
estimate of the types discussed in, e.g., \cite{Zurek,topB} to be important for our analysis.}, 
as there is a natural topological duality between the bubbles and the defect sites. These topologically-stable 
defects are the Skyrmions of the chiral theory, which we interpret as the electroweak baryons of the underlying 
strongly-coupled electroweak symmetry-breaking sector, for the reasons presented earlier.

By the time of percolation, the Universe would have filled with bubbles of the new vacuum with a
characteristic size $R$ given by
\begin{equation}
R \; \simeq \; \frac{3 S_0}{\Delta {\cal V}} ,
\label{radius}
\end{equation}
where $S_0 = \alpha_\epsilon \sqrt{B} V^3$ is the bubble action (\ref{bounce}), and $\Delta {\cal V} \simeq {\cal N} T_c^4$.
Using the estimate (\ref{epsilonvalue}) for $\alpha_\epsilon$, we find
\begin{equation}
R \; \simeq \; \frac{12 \alpha_\epsilon}{\sqrt{B} V} \; \simeq \; 24 \frac{\alpha_\epsilon}{m} \; \simeq \; \frac{1.35}{m} .
\label{size}
\end{equation}
It is clear that $R \ll H^{-1} \simeq (3/2\pi B)^{1/2} (M_P/V^2) \simeq (6/\pi)^{1/2} (M_P /m V)$.

If there is one electroweak baryon ${\cal B}$ per bubble, we may estimate their density to be
\begin{equation}
n_{\cal B} \; \sim \; \frac{1}{R^3} \; \sim \; 0.4 m^3 ,
\label{Bdensity}
\end{equation}
which can be compared with the entropy density
\begin{equation}
s \; \sim \; \rho^{3/4} \; \sim \; H^{3/2} M_P^{3/2} \; \sim \frac{m^{3/2} V^{3/2}}{8} ,
\label{sdensity}
\end{equation}
yielding an electroweak baryon-to-entropy ratio
\begin{equation}
\frac{n_{\cal B}}{s} \; \sim \; 3.2 \left(\frac{m}{V} \right)^{3/2} .
\label{Btos}
\end{equation}
This estimate is indisputably approximate, but it does suggest strongly that the baryon-to-entropy ratio is
likely to be {\it large immediately after percolation}.

However, this prediction of the Kibble-like mechanism for the production of electroweak baryons would have
been modified if equilibrium was restored after the transition. We now argue that this would have been the case 
following reheating after the electroweak transition, and that this would have led to a {\it much lower}
density of electroweak baryons.

The pseudo-dilaton would have decayed at percolation, and the residual vacuum
energy density thereby released would have led to reheating.
Since the pseudo-dilaton decay rate, which is $\propto \alpha_\chi m$ where $\alpha_\chi$ is some coupling strength,
is much larger than the Hubble rate after the transition, which is $\propto (m /M_P) m$, we expect that the
reheating temperature, $T_R$, would have been given by
\beq
\frac{\pi^2}{30} {\cal N} T_R^4 \simeq \frac{1}{4} B V^4 ,
\eeq
where ${\cal N} \sim {\cal O}(100)$ is the effective number of relativistic degrees of freedom
after the transition, implying that 
\beq
\label{reheat}
T_R \simeq 0.2 \sqrt{mV} .
\eeq
%Before concluding on the number density of pseudo-baryons, we must first check whether
%they would have equilibriated after bubble nucleation. 
Equilibrium would have been established following reheating
if the annihilation rate of electroweak baryons was at least as large as the Hubble expansion rate, i.e., if
\begin{equation}
\sigma v n \ga H \simeq \sqrt{\frac{8\pi^3 {\cal N}}{90}} \frac{T^{2}}{M_{Pl}} ,
\label{eqcondn1}
\end{equation}
where the annihilation cross section is related to the pseudo-dilaton mass: 
$\sigma v \simeq 1/4B\langle \chi \rangle^{2}$, $\langle \chi \rangle$ is its 
expectation value at reheating, and $n = (m_{\B}T/2 \pi)^{3/2} e^{-(m_{\B}/T)}$. One
expects that the effective electroweak baryon mass $m_\B \propto \langle \chi \rangle$.

Thus equilibrium would have been established if:
\begin{equation}
\frac{1}{4B \langle \chi \rangle^{2}} (\frac{m_{\B}}{2 \pi T})^{3/2} T 
e^{-(m_{\B}/T)}  \ga  \frac{\sqrt{{8 \pi^3 \cal N}/90}}{M_{P}} .
\label{eqcondn2}
\end{equation}
Defining $x \equiv (m_{\B}/T)$, this condition becomes
\begin{equation}
(32 \pi^6 {\cal N}/45)^{1/2} \frac{e^x}{\sqrt{x}} \la \frac{M_{P}m_{\B}}{m^{2}} ,
\label{eqcondn3}
\end{equation}
which may usefully be written in the form $x \ga x_F$, where the freeze-out temperature $x_F$ may
be expressed in the form
\begin{equation}
x_F \; = \; {\rm ln}(\frac{M_{P}}{V}) + 2{\rm ln}(\frac{V}{m}) + {\rm ln}(\frac{m_\B}{V}) - \frac{1}{2} {\rm ln}(32 \pi^6 {\cal N}/45) + \frac{1}{2}{\rm ln}(x_F) .
\label{x}
\end{equation}
We find the approximate value $x_F \simeq 40$, which is relatively insensitive to variations in $V/m \in (1, 10)$ 
and even less so to $m_\B/V$ if it is ${\cal O}(1)$, as expected,
corresponding to a freeze-out temperature
\begin{equation}
T_{F} \simeq \frac{m_{\B}}{40} .
\label{Tfreeze}
\end{equation}
Comparing with the estimate (\ref{reheat}) of the reheating temperature $T_R$, we see that $T_R > T_F$,
and hence that equilibrium is expected to have been achieved for $m/V \ga 0.015$ and $m_{\B} \simeq V$.

Once brought into equilibrium, the freeze-out density of electroweak baryons is
given simply by $n(T_F)$, namely
\beq
\frac{n_{\B}}{n_\gamma} \simeq \left(\frac{m_{\B}}{2 \pi T_F}\right)^{3/2} e^{-(m_{\B}/T_F)} .
\eeq
Using the estimate $T_F = m_{\B}/40$ gives
\beq
\eta_{\B} = \frac{n_{\B}}{n_\gamma} \simeq 10^{-16} ,
\label{ewbd}
\eeq
which is far below the Kibble estimate (\ref{Btos}) and in general agreement
with the estimate in \cite{CW}. A more detailed calculation would be
required to verify how accurately (\ref{ewbd}) is satisfied, but it is at least a useful lower limit.

For comparison, requiring the electroweak baryon density today to be $\sim 5$ times the
baryon density~\cite{WMAP}, taking the baryon-to-photon ratio $\eta \sim 6 \times 10^{-10}$ and
assuming a nominal electroweak pseudo-baryon mass $\sim 1$~TeV, we estimate the
electroweak baryon-to-photon ratio to be
\begin{equation}
\eta_{\cal \B} \; \sim \; 3 \times 10^{-12} .
\label{etaPB}
\end{equation}
The estimate (\ref{ewbd}) based on the freeze-out
abundance is much smaller, suggesting that
some mechanism for creating an electroweak pseudo-baryon asymmetry is needed if the electroweak 
pseudo-baryons are to provide all the dark matter~\cite{Nussinov}. However, we do not enter here into the details of any possible 
mechanism for generating such a electroweak pseudo-baryon asymmetry. Also, we note the possibility that
incomplete equilibration might have left the baryon density in the interesting range (\ref{etaPB}),
in which case an electroweak pseudo-baryon asymmetry might be unnecessary.

\section{Electroweak Baryons as Dark Matter}

In this section, we assume there is a cosmological electroweak baryon asymmetry
of the appropriate magnitude for the the electroweak baryons to acquire the
appropriate relic density $\Omega_{\cal B} h^2 \sim 0.1$~\cite{WMAP}. Since they interact
strongly at the TeV scale, one might wonder whether their self-interactions
would modify the standard WIMP scenario for dark matter. The self-interactions would be
dominated by exchanges of the lightest available particles in the spectrum,
namely the pseudo-dilaton and the massive electroweak gauge bosons.
These yield self-interacting cross sections that are ${\cal O}(1)$/(100~GeV)$^2$
at most, an upper bound strong enough for their self-interactions to have no
impact on the evolution of structures in the Universe.

We now study in more detail the direct interactions of electroweak baryons with
conventional matter, which would also be dominated by exchanges of the pseudo-dilaton
and the massive electroweak gauge bosons $W^\pm$ and $Z^0$. At this stage, we must 
distinguish the generic possibilities of bosonic and fermionic electroweak baryons. As 
discussed earlier, if the lowest-lying electroweak baryon state is bosonic, it is expected to have $I = J = 0$,
whereas if it is fermionic it is expected to have $I = J = 1/2$, like the conventional nucleons~\footnote{We
also note that, in the case of an Sp(N) gauge group, there would be no stable electroweak baryons,
as the electroweak Skyrmions would decay into bosons.}.
In the bosonic case, therefore, one need only consider pseudo-dilaton, $\gamma$ and $Z^0$
exchanges, whereas $W^\pm$ exchange should also be considered in the fermionic
case.

The questions also arise in the latter case which of the two (nearly) degenerate states
is lighter, and what is its electric charge. The answer to the latter question should clearly be that the
ground state is neutral, or the standard WIMP paradigm would fail~\cite{ZH2}. As discussed in~\cite{EMN},
two classes of diagrams are likely to contribute to the mass differences between the $I_3 = \pm 1/2$
fermionic partners: (a) electromagnetic `self-energy' diagrams and (b) photon-exchange `Coulomb potential'
diagrams. We expect these diagrams to have the following orders of magnitude:
\begin{equation}
{\cal O}(\frac{\alpha}{4 \pi}) V \times \left[ {\rm (a)} \Sigma_i q^2_i ~{\rm or}~ (\Sigma_i q_i)^2, {\rm (b)} \Sigma_{i \ne j} q_i q_j \right] .
\label{massdiff}
\end{equation}
It is plausible that, for any odd number of fermionic constituents of electroweak baryons, there are charge assignments
that ensure that the lighter of the two $I = 1/2$ states is electrically neutral. In the specific case of QCD, the
fact that the neutron is heavier than the proton is ascribed to the fact that $m_d > m_u$, whereas the electromagnetic
mass difference alone is calculated to yield $m_p > m_n$, so QCD actually supports the plausibility of this
expectation, despite the neutron being heavier! In any such scenario, 
the mass difference with the heavier member of the $I = 1/2$ doublet, as
given by (\ref{massdiff}), is expected to be ${\cal O}(1)$~GeV or more, as would also be given by scaling
up the electromagnetic contribution to the $p - n$ mass difference. We therefore expect that
the heavier (charged) electroweak baryon would $\beta$-decay with a lifetime ${\cal O}(10^{-11})$~s or less.

The electroweak baryon asymmetry might have either sign, so that the residual excess of electroweak baryons
might have electromagnetic charges $(0, +1)$ or $(-1, 0)$. Metastable charge $-1$ particles could affect Big Bang nucleosynthesis
via catalysis reactions that are absent for metastable charge $+1$ particles~\cite{Pospelov}. However, the above estimate indicates
that any charged electroweak baryons would decay before Big Bang nucleosynthesis, so it is unnecessary to
consider the corresponding constraints. Henceforward, we use the notation $({\cal N}, {\cal P})$ that is natural
for an $(0, +1)$ doublet, but our remarks apply equally to the $(-1, 0)$ case.

In general, the rate for weak-interaction ${\cal P} \leftrightarrow {\cal N}$ transitions at finite temperature is estimated to be
\begin{equation}
\Gamma_{{\cal P} \leftrightarrow {\cal N}} \; \sim \; G_F^2 T^5 ,
\label{weakrate}
\end{equation}
which may be compared with the Hubble expansion rate $H \sim \sqrt{g_* G_N} T^2$, where $g_*$ is the
effective number of relativistic degrees of freedom, which are assumed to dominate the cosmological
expansion. This comparison suggests that the weak interactions of the electroweak baryons freeze out at
\begin{equation}
T_F \; \sim \; \left( \frac{g_* G_N}{G_F^4} \right)^{1/6} \; \sim \; 1~{\rm MeV},
\label{freezeout}
\end{equation}
just like the weak interactions of conventional baryons. This would suggest a relic abundance of
${\cal P}$ relative to ${\cal N}$:
\begin{equation}
\frac{ n_{\cal P}}{n_{ \cal N}} \; \sim \; e^{- \Delta M/T_F} \; \sim \; e^{-10^3} ,
\label{PtoNthermal}
\end{equation}
which is completely negligible.

However, this argument neglects the possible formation of charged pseudo-nuclei that might be $\beta$-stable,
and be capable of trapping the otherwise-unstable ${\cal P}$ electroweak baryons. 
Any substantial relic abundance of such objects
would be incompatible with upper limits on possible charged relics from the Big Bang. In the case of
conventional Big-Bang nucleosynthesis, the formation of light nuclei is inhibited by photo-dissociation
until the density of photons above the nuclear photo-dissociation threshold falls below the density of
baryons, which occurs when the temperature $T \sim 0.1$~MeV, below the freeze-out temperature (\ref{freezeout}). 
In our electroweak baryon case, the density of photons above the photo-pseudo-dissociation threshold $\sim 2$~GeV falls
below that of the electroweak baryons when
\begin{equation}
e^{- 2~{\rm GeV}/T} \; \sim \; \eta_{\cal B} ,
\label{threshold}
\end{equation}
where $\eta_{\cal B}$ is given in (\ref{etaPB}). This condition is reached when the temperature falls
to $T \sim 75$~MeV, at which epoch
\begin{equation}
\frac{n_{\cal P}}{n_{\cal N}} \; \sim \; e^{- \Delta M/T} \; \sim \; 10^{-7} ,
\label{PoverN}
\end{equation}
assuming $m_{\cal P}-m_{\cal N} \sim 10^3 (m_n - m_p)$.
We conclude that the relative abundance of charged electroweak baryons ${\cal P}$ is suppressed strongly
below that of neutral electroweak baryons ${\cal N}$ well before the formation of pseudo-nuclei begins. 
However, the suppression (\ref{PoverN}) translates into an abundance relative to protons of
\begin{equation}
\frac{n_{\cal P}}{n_p} \; \sim \; 10^{-10} ,
\label{Poverp}
\end{equation}
which may not be sufficient by itself to respect the strong upper limits on the possible abundance of
charged relics from the Big Bang. Note that a even a slightly larger mass difference between
${\cal P}$ and ${\cal N}$, will greatly suppress the abundance of ${\cal P}$ to $p$. 

Detailed exploration of this question requires deeper studies of pseudo-nuclear physics and
Big-Bang pseudo-nucleosynthesis that go beyond the scope of this paper. However, we note that
if the abundance of charged relics turns out to be problematic, there are at least two escape
routes. One is that the electroweak baryons are bosons, in which case the lowest-lying state
has $I = J =0$, and the other is that the underlying gauge group is of Sp(N) type, in which case
electroweak baryons are unstable.

%%%%%%%%%%%%%%%%%%%%%%%%%%%%%%%%%%%%%%%%%%%%%%%%%%

%%%%%%%%%%%%%%%%%%%%%%%%%%%%%%%%%%%%%%%%%%%%%%%%%%

\section{Detection of Electroweak Baryonic Dark Matter}

The scattering of electroweak baryonic dark matter on conventional
matter is expected to be dominated by exchanges of the photon,
the neutral electroweak gauge boson $Z^0$ and the pseudo-dilaton $\chi$~\footnote{Scattering 
of a fermionic electroweak baryon via $W^\pm$ exchange
would entail inelastic charge-exchange: ${\cal N} p \to {\cal P} n$. Since
this requires an excitation energy ${\cal O}$(GeV), far exceeding the expected
kinetic energy of the relic dark matter particle, we neglect this contribution.}.

The photon-exchange contribution to technibaryon scattering on conventional matter
was first discussed in~\cite{BDT}. It was pointed out there that the electromagnetic
scattering of scalar electroweak baryons would proceed via their charge radius, which is quite
model-dependent. It was estimated that the rate would be relatively small (see also~\cite{FFS}),
and quite possibly below the reach of current experiments. On the other hand, it was also pointed out
in~\cite{BDT} that fermionic electroweak baryons would scatter via their magnetic-dipole 
moments, yielding considerably larger rates, of order 1 event/kg/day. A rate as large as
this has now been excluded by experiment~\cite{XENON100}, suggesting that fermionic electroweak baryons
could not constitute the astrophysical dark matter.

Turning to $Z^0$ exchange, it would contribute to the scattering of a fermionic electroweak baryon
via vertices that are both spin-independent and -dependent, but we do not discuss these in detail for
the reason given in the previous paragraph. On the other hand, $Z^0$-exchange would not contribute
to the scattering of a bosonic electroweak baryon, because it has $I = Y = 0$.

It remains to consider the pseudo-dilaton exchange contribution to the scattering
of bosonic electroweak baryons~\cite{FFS,HS}. Experiments usually
quote upper limits on the cross section on an individual nucleon $n$, neglecting the possible
difference between the proton and neutron. Neglecting the photon exchange contribution, the
cross section may be written in the form \cite{FFS}
\begin {equation}
\sigma_{\cal B} \; = \; \frac{m_R^2}{4\pi} \left[ \frac{f m_n M_{\cal B}}{V^2m^2}\right]^2 ,
\label{scalarsigma}
\end{equation}
where $m_R = M_{\cal B} m_n/(M_{\cal B} + m_n)$ is the reduced electroweak baryon mass.
In writing (\ref{scalarsigma}) we assume that all the electroweak baryon mass is due to the pseudo-dilaton
v.e.v. $V$, and the coefficient $f$ is defined and calculated as follows.

Following~\cite{fs}, we have
\begin{equation}
f \; = \; \sum_{q = u, d, s} f_q + \frac{6}{27} f_G ,
\label{definef}
\end{equation}
where for the light quarks $q = u, d, s$, we have:
\begin{eqnarray}
  \label{eqn:fNTu}
 f_u    & \ = \ & \frac{m_u B_u}{m_N}
    \; = \; \frac{2 \Sigma_{\pi N}}{m_N (1+\frac{m_d}{m_u})
                                 (1+\frac{B_d}{B_u})}
 \\
  \label{eqn:fNTd}
  f_d
    & \ = \ & \frac{m_d B_d}{m_N}
    \; = \; \frac{2 \Sigma_{\pi N}}{m_N (1+\frac{m_u}{m_d})
                                 (1+\frac{B_u}{B_d})}
    \\
  \label{eqn:fNTs}
 f_s
    & \ = \ & \frac{m_s B_s}{m_N}
    \; = \; \frac{(\frac{m_s}{m_d}) \Sigma_{\pi N} \, y}%
                   {m_N (1+\frac{m_u}{m_d})}
   \end{eqnarray}
and $f_G = 1 - \sum_{q = u, d, s} f_q$, where $\Sigma_{\pi N}$ is the $\pi$-nucleon
$\sigma$-term, $m_q B_q = \langle n|m_q \bar{q} q|n \rangle$, $y = 1 - \sigma_0/\Sigma_{\pi N}$, 
and octet baryon mass differences give $\sigma_0 = 36 \pm 7$~MeV. 
Inserting the following representative numerical values: $\Sigma_{\pi N} = 50$~MeV, $m_d/m_u = 1.81$, $B_d/B_u = 0.75$ and
$m_s/m_d = 18.9$, we estimate
\begin{equation}
f_u \; = \; 0.022, \; \; f_d \; = \; 0.029, \; \; f_s \; = \; 0.182, \; \; f_G \; = \; 0.767 ,
\label{fvalues}
\end{equation}
and hence $f \simeq 0.40$, though with considerable uncertainty, related in particular to the poorly-determined
experimental value of $\Sigma_{\pi N}$. Varying this between 36 and 72 MeV yields over an order of magnitude
range in $\sigma_{\cal B}$ for fixed values of the other parameters.

\begin{figure}[htb]
\begin{center}
\epsfig{file=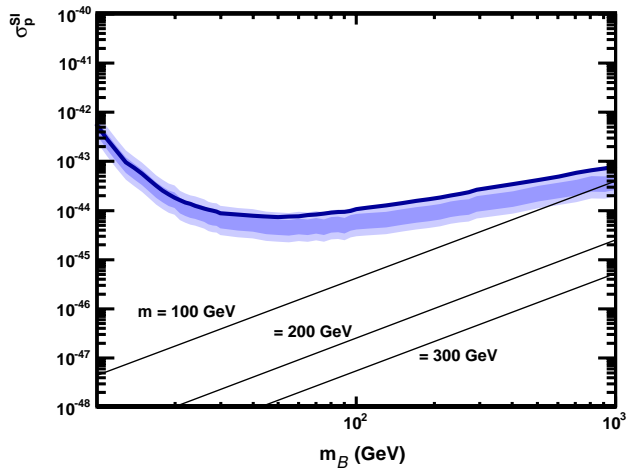,width=0.5\textwidth}
\caption{
\it The estimated cross section for the elastic scattering of a scalar electroweak boson ${\cal B}$ on a proton,
as a function of $m_{\cal B}$, assuming $V = 1$~TeV and the pseudo-dilaton mass $m = 100, 200$ and $300$~GeV. 
See the text for a discussion of the uncertainties in hadronic inputs to this calculation.
Also shown is the published upper limit of the XENON100 experiment
for dark matter particles masses $\le 1$~TeV (solid blue line) and its expected sensitivity (blue band)~\protect\cite{XENON100}.}
\label{fig:XENON}
\end{center}
\end{figure}

Numerically, assuming that the photon-exchange contribution is negligible, using $f = 0.4$
and choosing a representative electroweak baryon mass $M_{\cal B} = V = 1$~TeV
and the minimum dilaton mass $m = 100$~GeV, we find a cross section $\sigma_{\cal B} \simeq
4.4 \times 10^{-44}$~cm$^2$. This may be compared with the upper limit established by the XENON100
experiment on WIMP-nucleon scattering, which is $\sim 8 \times 10^{-44}$~cm$^2$ for a WIMP mass
of 1~TeV, the largest mass displayed in~\cite{XENON100}, as shown in Fig.~\ref{fig:XENON}. 
However, we note that the cross section (\ref{scalarsigma})
grows $\sim m_{\cal B}^2$, whereas the upper limit given in~\cite{XENON100} grows more slowly in the mass
range displayed. Extrapolating the published XENON100 result, as seen in Fig.~\ref{fig:XENON},
we estimate that the ${\cal B}p$ cross section may exceed the XENON100 sensitivity for
$m_{\cal B} \ga 2$~TeV, if $V$ = 1~TeV and $m= 100$~GeV. On the other hand, the calculated cross section decreases $\sim m^{-4}$
for larger pseudo-dilaton masses, as illustrated in Fig.~\ref{fig:XENON} by the cases $m = 200$ and 300~GeV. 
This discussion serves to emphasize, however, that there might be good prospects
for detecting electroweak dark matter with forthcoming experiments.

\section{Summary}

We have analyzed in this paper phenomenological and cosmological
aspects of what may be the best-defined extension of the Standard Model
in the context of composite models with new, nearly-conformal strong
interactions around the TeV scale. In this model there is a single additional 
light degree of freedom, in addition to the known Standard Model particles, 
a pseudo-dilaton. This particle would have couplings similar to those of a
Standard Model Higgs boson, except for a universal overall suppression and
the possibility that the loop-generated decays to $\gamma \gamma $ and $gg$
might be modified by additional particles with electric charges and/or QCD
interactions.
We have used the upper limits on Higgs production at the LHC~\cite{LHCH}
to constrain the pseudo-dilaton coupling suppression factor for dilaton
masses $m > 140$~GeV, and have also discussed the constraints coming
from precision electroweak data.

We have also discussed the cosmology of such a model, including the
possible nature of the electroweak phase transition, which we expect to
have been first-order except possibly for large $m$. This re-opens the
possibility of generating the cosmological baryon asymmetry during the
electroweak phase transition, a possibility we did not develop here. On the other hand, we do
not expect that an enormous growth in entropy would have been engendered
during the transition, so generating the cosmological baryon asymmetry
(and possibly the cosmological electroweak baryon asymmetry) before the
electroweak phase transition remains a possibility.

We identify electroweak baryons with the Skyrmion soliton solutions
of the low-energy effective electroweak theory, much as conventional
baryons may be described as Skyrmions in the low-energy effective
chiral SU(2) $\times$ SU(2) or SU(3) $\times$ SU(3) description of QCD.
In ignorance of the strong dynamics underlying electroweak symmetry
breaking, the electroweak baryons may be a doublet of $I = J = 1/2$ fermions or
a singlet $I = J = 0$ boson, or even unstable if the underlying gauge theory is
based on an SP(N) group. However, there are two potential problems
for the fermionic case: first, even if the charged member of the $I = J = 1/2$ doublet
is heavier than the neutral one, and hence unstable, some of its cosmological
abundance might have become `sequestered' in problematic charged stable `electroweak 
nuclei'; secondly, the magnetic-moment scattering of neutral fermionic electroweak baryons
is estimated to exceed the current upper limit~\cite{XENON100}. On the other hand,
this and other experiments are just at the verge of sensitive to scalar electroweak baryonic
dark matter.

The LHC experiments will presumably soon tell us whether there is a Standard Model
Higgs-like boson. If so, it would probably weigh less than about 130~GeV and our hunch is
that it would probably be accompanied by supersymmetric particles. On the other hand,
if it does {\it not} exist, one should not assume that the correct theory of electroweak
symmetry breaking is Higgsless. There may still exist a Higgs-like boson but with
suppressed coupling: a `less-Higgs' theory. In the event that a Standard Model-like
Higgs boson is not discovered soon with a mass below $\sim 140$~GeV,
the LHC experiments should push the upper limits on a `less-Higgs' boson down
relentlessly. Such a `less-Higgs' boson would necessarily be accompanied
by massive states, needed to unitarize $WW$ scattering, and its discovery would be almost
as exciting as proving that electroweak symmetry breaking is Higgsless.\\

{\bf Note added:}
While finishing this paper, three related papers have appeared. 
Constraints on a radion model have been considered recently in~\cite{deSR}.
Refs.~\cite{BIK,CGL} derive collider constraints on a model with a
dilaton coupled to gluons via the full QCD $\beta$ function for light quarks.
Because the light-quark contribution to the $\beta$ function is considerably larger than the top-quark contribution in
the model we study in our Section~2.3 and Fig.~1, the lower limit derived in~\cite{BIK,CGL} on their analogue of our 
dilaton v.e.v. $V$ is considerably stronger than in the model considered here. 

\section*{Acknowledgements}

J.E. thanks Marek Karliner for discussions.
The work of J.E. and K.A.O. was supported partly by the London
Centre for Terauniverse Studies (LCTS), using funding from the European
Research Council via the Advanced Investigator Grant 267352. 
The work of K.A.O. was also supported in part
by DOE grant DE--FG02--94ER--40823 at the University of Minnesota.
The work of B.A.C. was supported in part by a grant from the 
Natural Sciences and Engineering Research Council of Canada.
The authors thank the CERN Theory Division for its hospitality.

%%%%%%%%%%%%%%%%%%%%%%%%%%%%%%%%%%%%%%%%%%%%%%%%%%%%%%%%%

%%%%%%%%%%%%%%%%%%%%%%%%%%%%%%%%%%%%%%%%%%%%%%%%%%%%%%%%%

%%%%%%%%%%%%%%%%%%%%%%%%%%%%%%%%%%%%%%%%%%%%%%%%%%%%%%%%%

%%%%%%%%%%%%%%%%%%%%%%%%%%%%%%%%%%%%%%%%%%%%%%%%%%%%%%%%%

%%%%%%%%%%%%%%%%%%%%%%%%%%%%%%%%%%%%%%%%%%%%%%%%%%%%%%%%%

\end{document}